\documentclass[aps,prl,twocolumn,showpacs,psfig,superscriptaddress,longbibliography,floatfix]{revtex4-2}
\usepackage{epsfig}
\usepackage{graphics}
\usepackage{amsfonts}
\usepackage{mathrsfs}
\usepackage{amsmath}
\usepackage{color}
\usepackage{natbib}
\usepackage{textcomp}
\usepackage{graphicx}
\usepackage{bm}
\usepackage{amssymb}
\usepackage{ulem}
\usepackage{xspace}
\usepackage{epstopdf}
\usepackage{dcolumn}
\usepackage{longtable}
\usepackage{subfigure}
\usepackage{multirow}
\usepackage[colorlinks=true, pdfstartview=FitV, linkcolor=blue, citecolor=blue, urlcolor=blue]{hyperref}

\makeatletter

\newcommand{\Rmnum}[1]{\expandafter\@slowromancap\romannumeral #1@}
\makeatother
\newcommand{\Sc}{ScV$_6$Sn$_6$ }

\begin{document}

\title{Untangle charge-order dependent bulk states from surface effects in a topological kagome metal \Sc}
 
 \author{Zi-Jia Cheng}
 \email{zijiac@princeton.edu}
 \affiliation{Laboratory for Topological Quantum Matter and Advanced Spectroscopy (B7), Department of Physics, Princeton University, Princeton, New Jersey 08544, USA}

 \author{Sen Shao}
\affiliation {Division of Physics and Applied Physics, School of Physical and Mathematical Sciences, Nanyang Technological University, 21 Nanyang Link, 637371, Singapore}

\author{Byunghoon Kim}
\affiliation {Laboratory for Topological Quantum Matter and Advanced Spectroscopy (B7), Department of Physics, Princeton University, Princeton, New Jersey 08544, USA}

 \author{Tyler A. Cochran}
 \affiliation {Laboratory for Topological Quantum Matter and Advanced Spectroscopy (B7), Department of Physics, Princeton University, Princeton, New Jersey 08544, USA}

 \author {Xian P. Yang}
 \affiliation {Laboratory for Topological Quantum Matter and Advanced Spectroscopy (B7), Department of Physics, Princeton University, Princeton, New Jersey 08544, USA}

 \author {Changjiang Yi}
 \affiliation {Max Planck Institute for Chemical Physics of Solids,Nöthnitzer Straße 40, Dresden, 01187, Germany}
 
\author{Yu-Xiao Jiang}
\affiliation {Laboratory for Topological Quantum Matter and Advanced Spectroscopy (B7), Department of Physics, Princeton University, Princeton, New Jersey 08544, USA}

\author{Junyi Zhang}
\affiliation {Institute for Quantum Matter and Department of Physics and Astronomy, Johns Hopkins University, Baltimore, Maryland 21218, USA}

\author{Md Shafayat Hossain}
\affiliation {Laboratory for Topological Quantum Matter and Advanced Spectroscopy (B7), Department of Physics, Princeton University, Princeton, New Jersey 08544, USA}

 \author{Subhajit Roychowdhury}
 \affiliation{Max Planck Institute for Chemical Physics of Solids,Nöthnitzer Straße 40, Dresden, 01187, Germany}

 \author{Turgut Yilmaz}
 \affiliation{National Synchrotron Light Source II, Brookhaven National Laboratory, Upton, New York 11973, USA}

 \author{Elio Vescovo}
 \affiliation{National Synchrotron Light Source II, Brookhaven National Laboratory, Upton, New York 11973, USA}
 
 \author{Alexei Fedorov}
 \affiliation{Advanced Light Source, Lawrence Berkeley National Laboratory, Berkeley, CA 94720, USA}

 \author{Shekhar Chandra}
 \affiliation{Max Planck Institute for Chemical Physics of Solids,Nöthnitzer Straße 40, Dresden, 01187, Germany}

 \author{Claudia Felser}
 \affiliation{Max Planck Institute for Chemical Physics of Solids,Nöthnitzer Straße 40, Dresden, 01187, Germany}

 \author{Guoqing Chang}
 \email{guoqing.chang@ntu.edu.sg}
 \affiliation{Division of Physics and Applied Physics, School of Physical and Mathematical Sciences, Nanyang Technological University, 21 Nanyang Link, 637371, Singapore}

 \author{M.Zahid Hasan}
 \email{mzhasan@princeton.edu}
 \affiliation {Laboratory for Topological Quantum Matter and Advanced Spectroscopy (B7), Department of Physics, Princeton University, Princeton, New Jersey 08544, USA}

\begin{abstract}
Kagome metals with charge density wave (CDW) order exhibit a broad spectrum of intriguing quantum phenomena. The recent discovery of the novel kagome CDW compound \Sc has spurred significant interest. However, understanding the interplay between CDW and the bulk electronic structure has been obscured by a profusion of surface states and terminations in this quantum material. Here, we employ photoemission spectroscopy and potassium dosing to elucidate the complete bulk band structure of \Sc, revealing multiple van Hove singularities near the Fermi level. We surprisingly discover a robust spin-polarized topological Dirac surface resonance state at the M point within the two-fold van Hove singularities. Assisted by the first-principle calculations, the temperature dependence of the $k_z$- resolved ARPES spectrum provides unequivocal evidence for the proposed $\sqrt{3}$$\times$$\sqrt{3}$$\times3$ charge order over other candidates. Our work not only enhances the understanding of the CDW-dependent bulk and surface states in \Sc but also establishes an essential foundation for potential manipulation of the CDW order in kagome materials.

\end{abstract}
\maketitle

\section{I. Introduction}
Charge density wave (CDW), a collective quantum phenomenon wherein the electron density undergoes a periodic modulation, have long been extensively investigated in condensed matter physics\cite{wilson1975charge,gruner1988dynamics,mcmillan1976theory,carpinelli1996direct}. Arising from electron-lattice or electron-electron interactions, CDWs can compete with or coexist alongside other unconventional quantum states and offer a window into the complex behavior of correlated electrons\cite{chang2012direct,yu2021unusual,chen2016charge}. The emergence of CDWs in kagome metals has infused a fresh perspective into this field\cite{ortiz2019new,teng2022discovery}. Kagome metals, characterized by the distinctive frustrated lattice, furnish a valuable platform for exploring the interplay between topology, strong electronic correlation, and lattice dynamics\cite{yin2020quantum,ye2018massive,cheng2023visualization,yin2020fermion,yang2020giant}. Previous studies have demonstrated that a plethora of novel quantum phases, encompassing unconventional superconductivity and spontaneous time-reversal symmetry breaking, can arise in the kagome charge order materials \cite{jiang2021unconventional,mielke2022time,neupert2022charge,ortiz2020cs,zhao2021cascade, chen2021roton}. Understanding the genesis and impact of the CDWs within kagome metals constitutes an exciting frontier, imbued with immense potential to provide deep insights into uncharted phases and complex electron-lattice dynamics.

\begin{figure*}[!htb]
	
	\begin{center}
		\includegraphics[scale=0.7, angle=0]{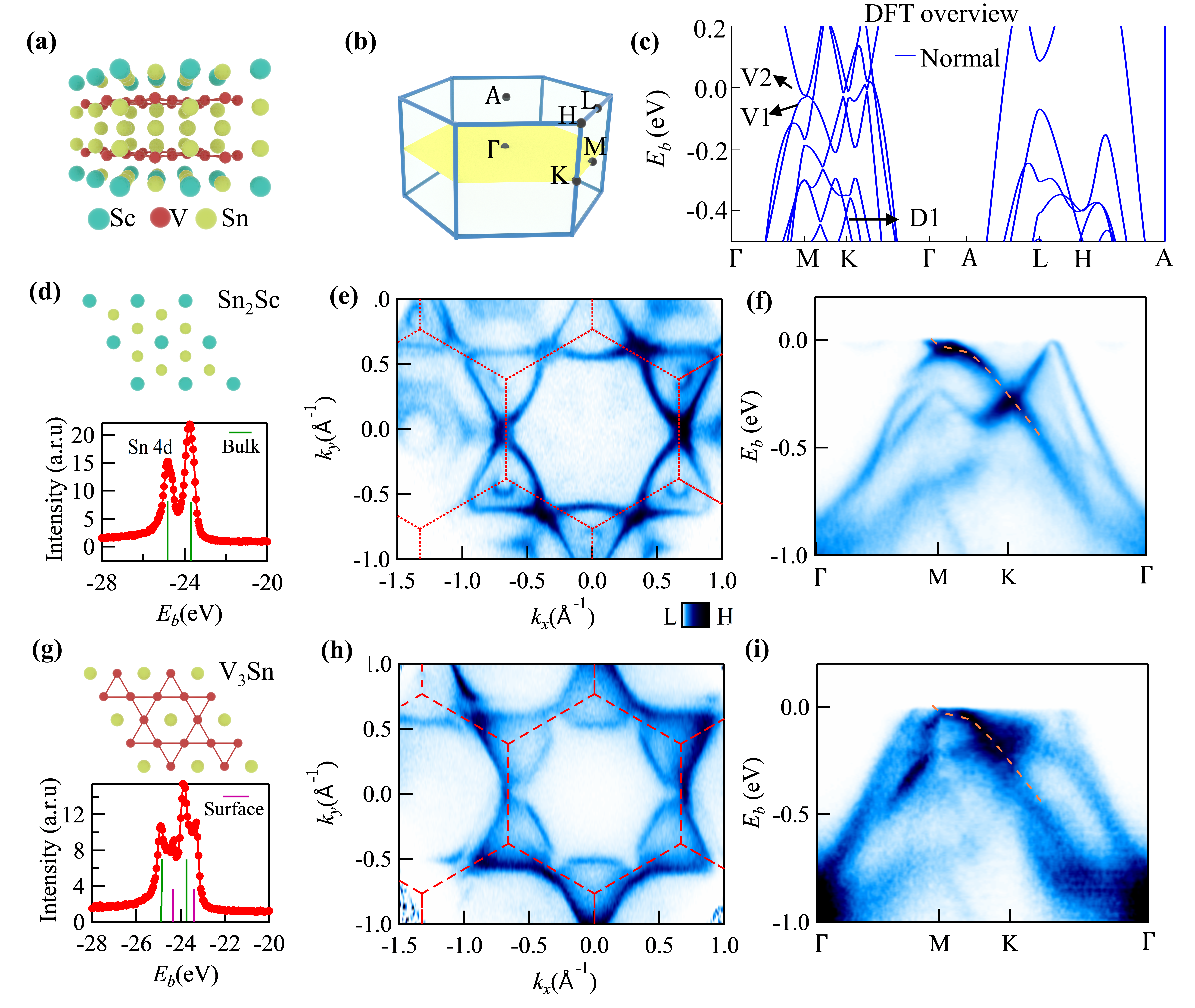}
	\end{center}
	\caption{Termination-dependent band structure in \Sc . (a) Side view of the crystal structure of \Sc in the normal phase. (b) Brillouin zone (BZ) and high symmetry points. (c) Theoretical band structure of normal phase of \Sc. The two types of van Hove singularities (V1 and V2) and Dirac cone (D1) are marked with arrows, respectively. (d) Core-level spectrum of Sn 4d orbitals on the Sc$_2$Sn layer. The corresponding Fermi surface map, measured with 80 eV photon energy and at 10K, is shown in (e). (f) Energy-momentum cuts along high-symmetry lines. For visual clarity, a dashed line has been plotted at the V2 state. (g), (h) Same with (d-f) but at V$_3$Sn termination. A notable difference between (e, f) and (h, i) emphasizes the presence of surface states}
	\label{fig1}
\end{figure*}

Recently, \Sc has been recognized as a novel kagome metal with CDW ground state at low temperatures and sparked significant interest\cite{arachchige2022charge}. The DFT calculations reveal the existence of two-fold van Hove singularities (VHS) near the Fermi level ($E_f$) and a Dirac point [Fig.~\ref{fig1}(c)], characteristic of kagome metals. However, unlike the prototypical quasi-2D kagome charge order compounds AV$_3$Sb$_5$ (A= K, Cs, Rb)\cite{jiang2021unconventional,wang2021electronic} and FeGe\cite{teng2022discovery,teng2023magnetism,shao2023intertwining}, \Sc hosts two layers of kagome lattices within a unit cell and exhibits a distinct three-dimensional modulation vector in bulk\cite{arachchige2022charge}. Despite being subjected to intensive transport and spectroscopic scrutiny, the nature of the CDW and its interplay with bulk electronic structure and topology remain intensively contested\cite{di2023flat,tan2023abundant,kang2023emergence,lee2023nature,hu2023kagome,hu2023phonon}. The investigation of bulk electronic structure is greatly obfuscated by the presence of numerous surface states on the multiple terminations, as well as the inherent three-dimension nature of the CDW order. Thus, determining the actual bulk band structure and unearthing the signature of CDW with $k_z$ resolution has become a crucial step towards unraveling the origin and effect of the phase transition, laying the groundwork for its future applications.

In this letter, we successfully disentangle the surface and bulk states of \Sc by combining high-resolution angle-resolved photoemission spectroscopy (ARPES) and K dosing. We further demonstrate the Z2 topology within the gap of the two-fold VHSs, through revealing the spin-momentum locking of the robust Dirac surface resonance state near the M point. With an enriched understanding of the intricate bulk band structure and DFT calculations, the measured ARPES spectrum across the CDW transition temperature provides unambiguous evidence for $\sqrt{3}$$\times$$\sqrt{3}$$\times$3 charge order and excludes other scenarios.

\section{II. Methods}

High-quality single crystals of \Sc were grown by the flux method. Starting elements Sc, V, and Sn with a molar ratio of 1: 10: 60 were loaded in an alumina crucible and then sealed in an evacuated silica tube. The tube was then slowly heated to 1373 K, dwelt for 10 hours, and then cooled down to 973 K over 400 hours. Silvery crystals with a hexagonal shape and a typical size of 2$\times$2$\times$1 mm$^3$ were yielded after centrifugation.

The VUV ARPES measurements were performed at the Electron Spectro-Microscopy (ESM) beamline, National Synchrotron Light Source II, Brookhaven national laboratory. The beam spot was less than 20 $\times$ 10 $\mu m^2$. The \Sc samples were cleaved at 10 K to expose a fresh surface. The energy resolution was better than 20 meV and the angle resolution was 0.2 degree. For the K dosing, we heat the K source with a 5.5 A current for 20 minutes. Several spectra were measured during the process until the energy-momentum cut didn't show further changes. The spin-ARPES measurement was performed at the beamline 10.0.1.2, Advanced Light Source, Lawrence Berkeley National Laboratory. The Soft X-ray ARPES results were obtained at the BL25SU beamline of SPring-8, with an energy resolution of 100 meV. Based on the photon-energy dependence measurement, an inner potential of 16 eV was used for determining the high symmetry energies. Throughout the text, we use the unit cell of the normal phase to determine the BZ. Multiple samples were measured, which yielded consistent results. All the data were measured at the base temperature except the temperature dependence measurement. 

We performed $\textit{ab initio}$ calculations in the framework of density-functional theory within the Perdew-Burke-Ernzerhof exchange-correlation functional \cite{perdew1996generalized}, as implemented in Vienna ab-initio Simulation Package (VASP) \cite{kresse1996efficient}. The all-electron projector augmented-wave method \cite{blochl1994projector} was adopted with 3$\textit{d}$$^{1}$4$\textit{s}$$^{2}$, 3$\textit{d}$$^{3}$4$\textit{s}$$^{2}$ and 5$\textit{s}$$^{2}$5$\textit{p}$$^{2}$ treated as valence electrons for Sc, V and Sn atoms, respectively. A cutoff energy of 400 eV and a dense k-point sampling of the Brillouin zone with a KSPACING parameter of 0.2 were used to ensure the enthalpy converged within 1 meV/atom. We adopted zero damping DFT-D3 van der Waals correction \cite{grimme2010consistent} in our calculations. The band topology was calculated from the parity, which was generated by vasp2trace \cite{vergniory2019complete}. The post-processing of band and unfolding band calculations was performed by the VASPKIT tool \cite{wang2021vaspkit}. The surface states calculations were carried out using the combination of Wannier90 \cite{Pizzi2020} and WannierTools \cite{WU2017} software packages.

\section{III. Results and Discussion}

 Similar to the other members of a versatile crystal family denoted by the formula ATm$_6$Sn$_6$ (A is rare earth or S-block element and Tm is transition metal)\cite{ma2021rare,cheng2023visualization}, \Sc crystallizes in space group $\textit{P}$6/$\textit{mmm}$, accommodating two A-A stacked V kagome lattices that are sandwiched by the Sn$_2$Sc layer and Sn layer [Fig.~\ref{fig1}(a)]. The weak bonding between V and Sc atoms establishes the natural cleavage plane at the juncture of these two layers\cite{hu2022tunable}. Hence, two terminations, labeled as Sn$_2$Sc termination and V$_3$Sn termination, are equally likely to appear in the experiment. In line with our expectations, we observed two distinct Sn-4d core levels (comprising 4d$_{\frac{3}{2}}$ and 4d$_{\frac{5}{2}}$) during scanning the micro-level beam across the sample surface, as depicted in Fig.~\ref{fig1}(d) and Fig.~\ref{fig1}(g). The presence of side peaks in Fig.~\ref{fig1}(g) implies a distinct chemical environment of Sn atom at the surface comparing to the bulk in this termination. The correspondence between the Sn core levels and terminations is currently under debate\cite{hu2022tunable,lee2023nature,di2023flat,peng2021realizing}. In this paper, we attribute the second type of core level to the polar environment of Sn atom at V$_3$Sn termination and assign the other one to the Sn$_2$Sc termination. The assignment finds support in the agreement between the termination-dependent ARPES results, as shown in Figs.~\ref{fig1}(e-f, h-i) and surface-projected DFT calculations [Fig. S1\cite{SI}]. Notably, the hexagonal Rashba surface state appearing in the second BZ [Fig.~\ref{fig1}(e)] is well reproduced in the corresponding surface's calculation [Fig. S1(b)\cite{SI}]. Although the bulk Dirac point (D1) and its associated van Hove band may be potentially identified by comparing the energy-momentum cuts along high symmetry lines [Figs.~\ref{fig1}(f, i)] of two terminations, the profusion of discrepancies in the spectrum poses a significant challenge in identifying the genuine bulk state and further examining of the effect of bulk charge density wave. 

\begin{figure}[!tb]
	\centering
	\includegraphics[scale=0.63,angle=0]{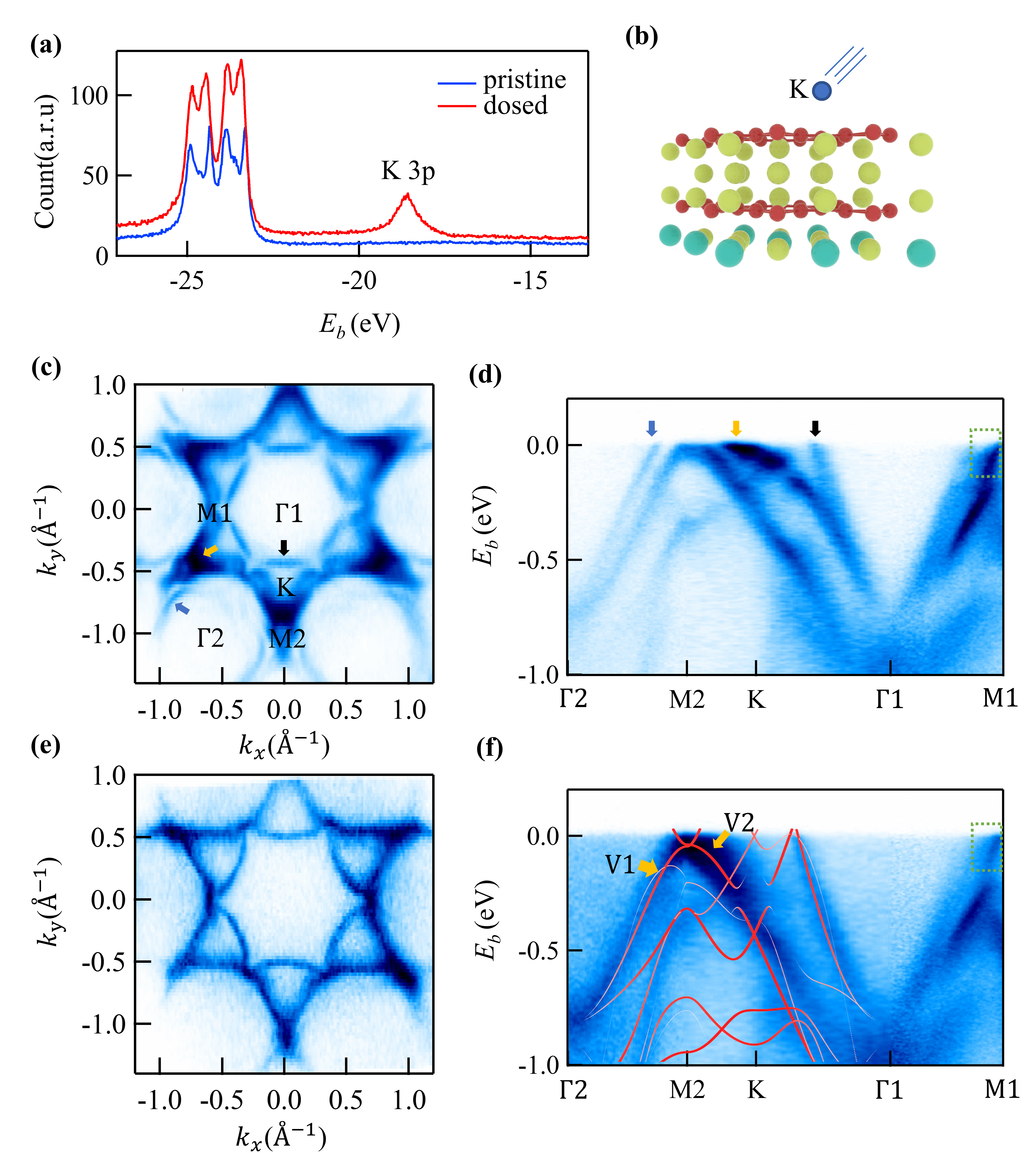}\\
	\caption{Unravel the bulk band structure with K dosing. (a) Core level spectrum of Sn 4d state and K 3p state before and after the K dosing. Throughout the dosing process, the number of Sn 4d peaks remains unchanged, confirming measurements taken at the V termination. (b) A schematic provides a visual representation of the K dosing at the V$_3$Sn termination. (c), (e) The Fermi surface map acquired on the pristine surface (c) and the surface after dosing (e). 129 eV light with linear horizontal polarization was used for collecting the data. (d), (f) Energy-momentum cuts along high symmetry directions, derived respectively from the maps in (c) and (e). The V d-orbital-resolved DFT calculation is overlaid and shows remarkable consistency with the ARPES features near the Fermi level. }\label{fig2}
\end{figure}

\begin{figure*}[!htb]
	\begin{center}
        \includegraphics[width=0.9\textwidth,angle=0]{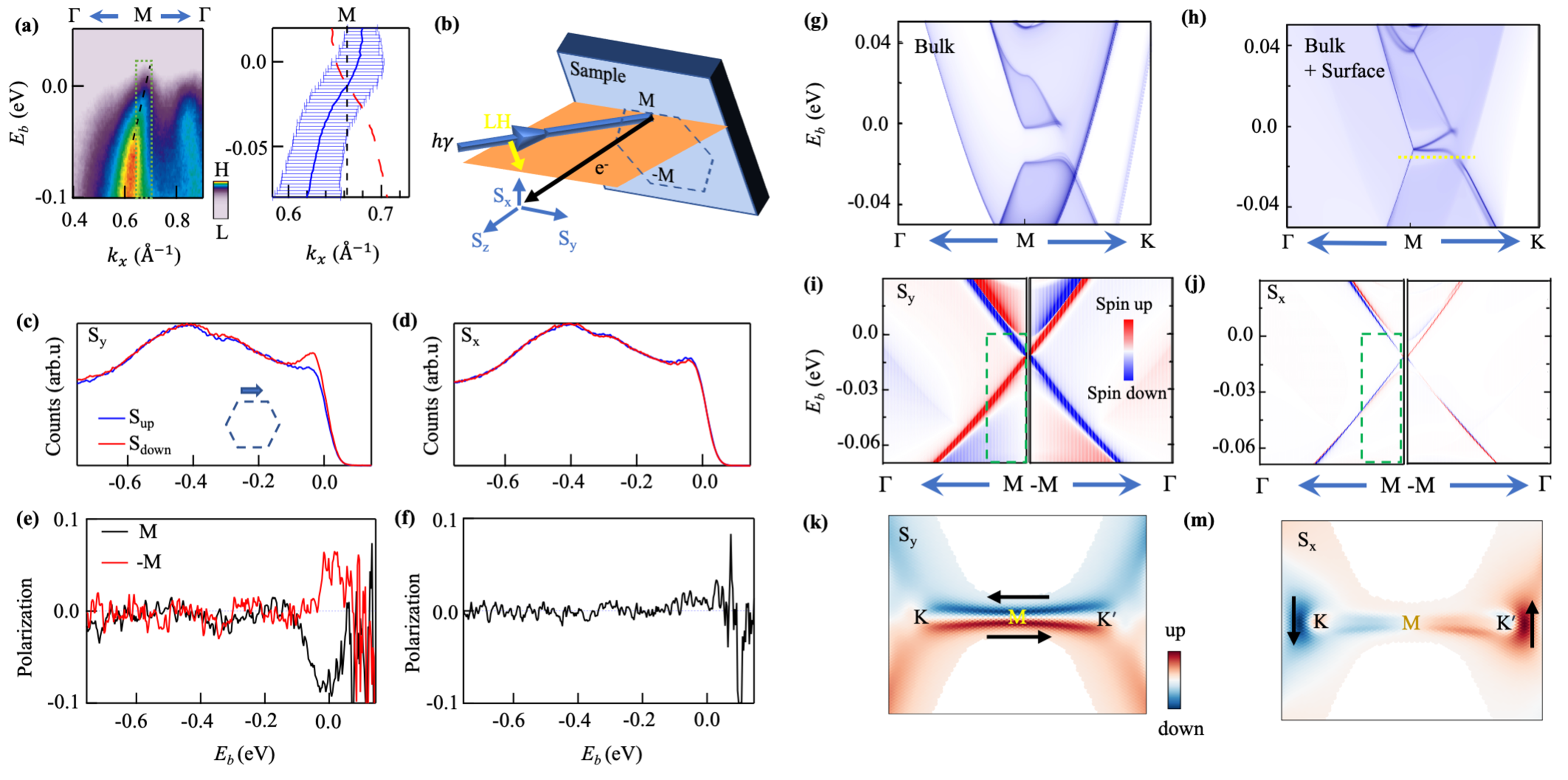}
	\end{center}
	\caption{Spin polarization of the Dirac surface resonance state at the M point. (a) Left: Energy - momentum cut along $\Gamma$-M direction, acquired with 50 eV photons. A green rectangle accentuates the detected range of spin EDCs. Right: Extracted and symmetrized dispersion of the S1 state, revealing the full linear crossing around the M point. (b) The geometry of the spin-resolved ARPES measurement. (c), (d) Spin-resolved energy-dispersion curves (EDCs) for the S$_y$ direction (c) and the S$_x$ directions (d). The inset in (c) figures depicts the relative orientation of the spin polarization in relation to the BZ. (e), (f) The binding energy dependence of net spin polarization along S$_y$ and S$_x$. Data measured at the opposite M point (-M) were plotted and revealed the opposite spin polarization. (g) Surface-projected bulk band structure along $\Gamma$ - M - K path of the surface BZ. (h) Calculated surface spectrum at V$_3$Sn termination, in which a singularly-degenerated Dirac state emerges between the two VHSs. (i), (j) Computed spin polarization of the bands near the M point. The green dashed rectangles in (i) and (j) represent the effective detection regions of spin EDCs in (c) and (d), respectively. (k),(m) In-plane spin polarization distribution at binding energy -18 meV (green dashed line in (h)).}\label{fig3}
\end{figure*}

To eradicate the surface states induced by the dangling bond of V$_3$Sn termination, we dosed the V surface with alkaline metal and examined the consequent change in the ARPES spectrum at k$_z$$=$ 0 of the bulk BZ [Figs.~\ref{fig2}(a,b)]. Figures ~\ref{fig2}(c-f) summarize our findings, in which data across multiple BZs and high symmetry lines are incorporated to compensate for the matrix element and sublattice interference effect. Interestingly, after moderate K absorption, the spectrum weight of three states near the $E_f$ [marked with colored arrows in Figs.~\ref{fig2}(c-d)] diminishes, indicative of their surface state nature. Notice that the state pointed by the yellow arrow was previously identified as a CDW-driven surface state of the kagome termination\cite{kang2023emergence}, and its absence under K dosing indicates the instability of the CDW on the surface. In contrast, bands V1 and V2 exhibit a rigid band shift of approximately -20 meV due to electron doping and form a closed Star-of-David shape pattern at the $E_f$. As the intensity of a single kagome-driven Dirac band exhibits BZ-selected behavior due to sublattice interference, such a pattern is consistent with the existence of two distinct van Hove singularities. A thorough analysis of the ARPES spectrum reveals the two van Hove singularities, initially nearly degenerate at $E_b =$ -0.02 eV pre-dosing, shift downwards to $E_b =$ -0.042 eV post-dosing. For the first time of \Sc, the orbital-decomposed bulk DFT calculation shows remarkable consistency with the post-dosing spectrum along the $\Gamma$2-M2-K-$\Gamma$1 path, which not only substantiates the assignment of bulk and surface states but also suggests the V1 and V2 VHS originate from $d_{x^2-y^2}$ and $d_{z^2}$ orbitals, respectively (see orbital-resolved calculation in Fig. S2\cite{SI}). The quantitative consistency of the band structure also points to the weak electronic correlation of the systems, thereby excluding the possibility of the strong correlation as the driving force of the CDW order\cite{sobota2021angle}.

\begin{figure*}[!htb]
	
	\begin{center}
		\includegraphics[width=0.9\textwidth]{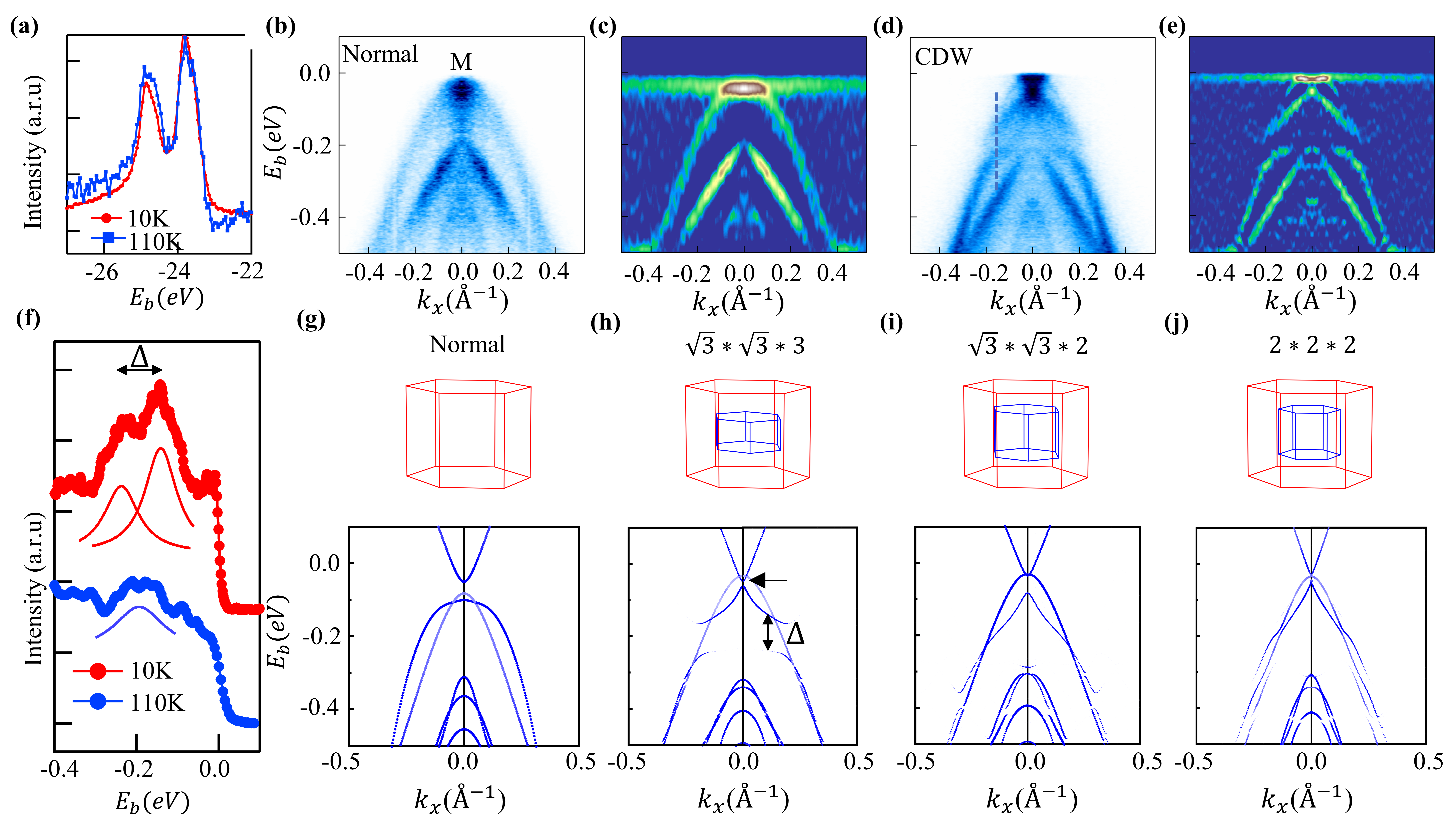}
	\end{center}
	\caption{Direct observation of band folding effect of $\sqrt{3}$$\times$$\sqrt{3}$$\times$3 charge order. (a) Core level spectroscopy of Sn 4d core level measured at both 10 K and 110 K. (b) Cut along $\Gamma$ - M - $\Gamma$ direction, obtained with 80 eV photon. Its second derivative is shown in (c). (d), (e) Similar cut with (b-c) but measured below the CDW transition temperature, showing a distinct gap feature below the Fermi level. (f) EDCs cross the gap feature, which are extracted from the dashed line in (d) and the same momentum location in (b). (g), (f) Unfolded band structure in various configurations: the normal phase (g), $\sqrt{3}$$\times$$\sqrt{3}$$\times$3 order (h),$\sqrt{3}$$\times$$\sqrt{3}$$\times$2 order (i) and 2$\times$2$\times$2 charge order (j). The corresponding BZs (red for the normal phase and blue for the CDW phases) are also presented above the calculations. The consistency between (h) and (d) confirms the type of order.}\label{fig4}
 \end{figure*}

However, a noteworthy state [referred to as S1, circled by the green rectangle in Figs.~\ref{fig2}(d,f)] appearing at the M point of the first BZ shows no observable change with K dosing while doesn't align with any bulk state from the DFT calculations. As shown in Fig.~\ref{fig3}(a), its linear dispersion closely resembles the Dirac surface state in the topological insulators, and the energy position of the crossing point coincides with the two-fold VHSs. To further investigate the nature of this robust state, we carried out spin-resolved ARPES (SR-ARPES) measurements on the same state at 10 K with 50 eV photons [Figs.~\ref{fig3}(a),(b)]. By measuring the spin-resolved energy distribution curve (EDC) around M point [green box in Figs.~\ref{fig3}(a),(g).(h)], we observed a clear difference between the spin-up and spin-down EDC curves at $E_f$ for S$_y$, while negligible difference in S$_x$ [shown in Figs.~\ref{fig3}(c-f)]. This result directly demonstrates the direction-selective in-plane spin polarization of the S1 state. Additionally, the polarization flips the sign when measuring at the opposite M point (-M), which indicates the preserved macroscopic time-reversal symmetry within the CDW phase. The consistency of these observations across different photon energies further negates the final state effect (see Fig. S3 in Supplemental Material\cite{SI}). The existence of spin polarization serves as direct evidence for the surface state nature of S1, considering the bulk is inversion symmetric.

We computed the surface-projected bulk-band structure and surface states from the first principles, the results of which are shown in Fig.~\ref{fig3}(g-h).  The calculations confirm the existence of a linear-dispersing surface resonant band between the two VHS bands at the M point on the V$_3$Sn surface, in agreement with the ARPES observations. Due to the k$_z$ dispersion of the bulk states, the Dirac state overlaps with the bulk-band continuum, and the allowable surface-bulk hybridization may explain its robustness under K dosing. Spin-resolved calculations further demonstrate a pronounced S$_y$ polarization and negligible S$_x$ polarization in the S1 state along $\Gamma$ – M direction (Fig. 3(i, j)), which fully align with the spin-ARPES observations. We note the vanishing S$_x$ polarization is enforced by the M$_y$ symmetry on the (001) surface. The presence of spin-polarized Dirac surface resonance is a direct consequence of the non-trivial Z2 topology of the V1 band\cite{fu2007topological} (see details in Fig.S4 of Supplement Material\cite{SI}) and therefore is topologically protected. Interestingly, despite the strong anisotropy in the band dispersion, the topological surface resonance exhibits a spin-momentum locking feature (Fig.~\ref{fig3}(k-m)), i.e., the spin direction follows the momentum trajectory on a constant energy surface. As the S1 state crosses the Fermi level, its highly anisotropic spin-momentum locking characteristic emerges as a potent mechanism for generating direction-sensitive spin-orbit torque\cite{shao2021roadmap}, making it a valuable component for future topological spintronics.

Prior research has demonstrated that Scandium and Tin atoms undergo profound out-of-plane displacement in the CDW phase, whereas the Vanadium atom's movement is minimal \cite{arachchige2022charge}. Therefore, the bands associated with the Sc and Sn orbitals are anticipated to exhibit greater changes during the phase transition. To directly probe the p orbitals of Sn atoms, we now turn to the Sn$_2$Sc termination and inspect the CDW's impact on the band structure. We measured the same energy-momentum cut along $\Gamma-M-\Gamma$ at 80 eV, corresponding to $k_z \approx 0.3 \pi/c$ (near the BZ boundary of $\sqrt{3}$$\times$$\sqrt{3}$$\times$3 order), below and above CDW transition temperature $T_c =$ 92 K [Fig.~\ref{fig4}(a)]. The results and their second derivatives are displayed in Figs.~\ref{fig4}(b-e). Above $T_c$, the main feature near the $E_f$ is a quadratic hole-like band with a band top $E_b =$ -0.03 eV. In the CDW phase, the hole-like band breaks into two distinct bands: namely, a Dirac-like band with crossing energy -0.03 eV and a deeper hole-like band. The breaking point locates at $\approx$ -0.2 eV and $\approx$ -0.2 \AA$^{-1}$, resulting in a sizable gap $\Delta$ of 0.1 eV [Fig.~\ref{fig4}(f)]. As the gap is not positioned at $E_f$ and the spectrum weight diminishes away from the gap, it mainly comes from the intrinsic electronic hybridization between backfolded bands, instead of diffraction of the final states\cite{watson2023spectral}. Meanwhile, no charge order gap at the Fermi level, which lowers the total electronic energy, is observed at this $k_z$. 

The unique band structure transformation at k$_z$ $\approx$ 0.3 $\pi$/c imposes strict constraints on the nature of the charge density wave. To elucidate this, we conducted DFT calculations for various possible charge orders within the systems, including 2$\times$2$\times$2 (space group $\textit{Immm}$), $\sqrt{3}$$\times$$\sqrt{3}$$\times$2 (space group $\textit{P}$6/$\textit{mmm}$) and $\sqrt{3}$$\times$$\sqrt{3}$$\times$3 (space group $\textit{R}$32) and present the corresponding energy-momentum cut of Sn 4p orbital contribution in Figs.~\ref{fig4}(h-j), respectively. We note that a band-unfolding procedure was utilized in the calculation for obtaining the spectrum weight in the first BZ of the normal phase and further facilitating the comparison\cite{medeiros2014effects}. Evidently, only the result of the $\sqrt{3}$$\times$$\sqrt{3}$$\times$3 order, which was previously determined via XRD, perfectly matches with the ARPES observation. Other charge orders, despite having lower or comparable ground state energies according to the calculation, fail to account for the valence band gap opening and thus are excluded based on the above argument. Meanwhile, the quadratic band feature can be well captured with the calculation of the normal phase [Fig.~\ref{fig4}(f)], further validating the DFT calculations (see more discussion in Figs. (S5,S6) of Supplemental Material\cite{SI}).

Interestingly, a recent experiment has observed a collapsed phonon mode with q = $(\frac{1}{3},\frac{1}{3},\frac{1}{2})$ below $T_c$ \cite{korshunov2023softening}. The absence of a band renormalization that corresponds to $\sqrt{3}$$\times$$\sqrt{3}$$\times$2 order, as demonstrated in this work, hints at the non-static nature of the phonon softening. Concurrently, we observed a large $k_z$ dispersion of the VHS state V2, which is hole-like along $M-L$ with 0.4 eV out-of-plane bandwidth (see Fig. S7(c) in Supplemental Material\cite{SI}). In stark contrast to the quasi-2D dispersion of VHSs in CsV$_3$Sb$_5$ \cite{kang2022twofold}, the out-of-plane dispersion in \Sc is much larger than the bandwidth of the phonon mode and may impede the nesting conditions of the VHSs and attenuate the divergence of the density of states, despite the fact that the VHSs are near the Fermi level at $k_z$ = 0. Those factors are the key to realizing a charge density wave with in-plane Q vector 2$\times$2 \cite{rice1975new}. Thus, the comprehensive three-dimensional characteristics of the system must be considered when analyzing the mechanisms underlying the CDW and the absence of superconductivity in \Sc.

In summary, our study delivers a comprehensive and rigorous examination of the bulk and surface band structure of \Sc. The identification of the robust topological surface resonance state at M point and the observation of its spin polarization reveals the topology of the system's electronic structure. More importantly, our $k_z$-resolved ARPES spectrum, substantiated by DFT calculations, directly demonstrates the existence of $\sqrt{3}$$\times$$\sqrt{3}$$\times$3 charge order and the exclusion of other charge orders, offering key insights into the three-dimensional nature of the CDW phase. Our findings lay the foundation for the exploration and manipulation of CDW orders within the growing list of correlated kagome materials. 

\nocite{cheng2023visualization,perdew1996generalized,kresse1996efficient,blochl1994projector,kresse1996efficient,vergniory2019complete,wang2021vaspkit,Pizzi2020,WU2017}

\section*{Acknowledgements}

The authors thank Haoyu Hu for the fruitful discussions. Work at Princeton University and Princeton-led synchrotron-based ARPES measurements were supported by the U.S. Department of Energy (DOE) under the Basic Energy Sciences program (grant no. DOE/BES DE-FG-02-05ER46200). Theoretical works at Princeton University were supported by the Gordon and Betty Moore Foundation (GBMF9461; M.Z.H.). This research used Beamline 21-ID-1 (ESM-ARPES) of the National Synchrotron Light Source II, a U.S. Department of Energy (DOE) Office of Science User Facility operated for the DOE Office of Science by Brookhaven National Laboratory under Contract No. DE-SC0012704. This research also used resources of the Advanced Light Source, which is a DOE Office of Science User Facility under Contract No. DE-AC02-05CH11231. Synchrotron radiation experiments were performed at the BL25SU of SPring-8 with the approval of the Japan Synchrotron Radiation Research Institute (JASRI) (Proposal No. 2023A1611). Part of this work was financially supported by the Deutsche Forschungsgemeinschaft (DFG) under SFB1143 (project no. 247310070) and the Würzburg-Dresden Cluster of Excellence on Complexity and Topology in Quantum Matter—ct.qmat (EXC 2147, project no. 390858490). Work at Nanyang Technological University was supported by the National Research Foundation, Singapore under its Fellowship Award (NRF-NRFF13-2021-0010) and the Nanyang Assistant Professorship grant. S.R. thanks the Alexander von Humboldt Foundation for a fellowship.

Z.-J. C., S. S., and B. K. contributed equally to this work.

\bibliography{ref.bib}

\end{document}